\documentclass[journal]{IEEEtran}
\usepackage{amsmath,amsfonts}
\usepackage{algorithmic}
\usepackage{algorithm}
\usepackage{array}
\usepackage[caption=false,font=normalsize,labelfont=sf,textfont=sf]{subfig}
\usepackage{textcomp}
\usepackage{stfloats}
\usepackage{url}
\usepackage{verbatim}
\usepackage{graphicx}
\usepackage{cite}
\usepackage{amsmath}
\usepackage{xcolor}

\begin{document}

\title{Orthogonal Spin Current Injected Magnetic Tunnel Junction for Convolutional Neural Networks}

\author{\IEEEauthorblockN{Venkatesh Vadde$^{1}$,
Bhaskaran Muralidharan$^{1,}$\IEEEauthorrefmark{1},
Abhishek Sharma$^{2,}$\IEEEauthorrefmark{2}}

\IEEEauthorblockA{$^{1}$Department of Electrical Engineering, Indian Institute of Technology Bombay, Powai, Mumbai-400076, India}
\IEEEauthorblockA{$^{2}$Department of Electrical Engineering, Indian Institute of Technology Ropar, Rupnagar, Punjab-140001, India}

\thanks{\IEEEauthorrefmark{1}bm@ee.iitb.ac.in, \IEEEauthorrefmark{2}abhishek@iitrpr.ac.in}}
\maketitle

\begin{abstract}

%Conventional logic and memory design using spintronics employ stable nanomagnets with energy barriers of more than $40 kT$. In this work, we show that an unstable ferromagnet can be used in the hardware design of neuromorphic computing. 
We propose that a spin Hall effect driven magnetic tunnel junction device can be engineered to provide a continuous change in the resistance across it when injected with orthogonal spin currents. Using this concept, we develop a hybrid device-circuit simulation platform to design a network that realizes multiple functionalities of a convolutional neural network. At the atomistic level, we use the Keldysh non-equilibrium Green's function technique that is coupled self-consistently with the stochastic Landau-Lifshitz-Gilbert-Slonczewski equations, which in turn is coupled with the HSPICE circuit simulator. We demonstrate the simultaneous functionality of the proposed network to evaluate the rectified linear unit and max-pooling functionalities. We present a detailed power and error analysis of the designed network against the thermal stability factor of the free ferromagnets. Our results show that there exists a non-trivial power-error trade-off in the proposed network, which enables an energy-efficient network design based on unstable free ferromagnets with reliable outputs. The static power for the proposed ReLU circuit is $0.56\mu W$ and whereas the energy cost of a nine-input rectified linear unit-max-pooling network with an unstable free ferromagnet($\Delta=15$) is $3.4pJ$ in the worst-case scenario. We also rationalize the magnetization stability of the proposed device by analyzing the vanishing torque gradient points. 
%This work also lays the ground for the development of an integrated spintronics-CMOS hybrid simulation platform that unites an atomistic understanding of spin currents with functional circuit design. 

\end{abstract}

\section{Introduction}

Neuromorphic computing takes inspiration from biological brains to perform highly complex problems while consuming remarkably low energy \cite{markovic2020physics, schuman2017survey}. The brain performs in-memory computation and uses many low-precision calculations in parallel to perform a task. Meanwhile, modern computers are primarily based on the von Neumann architecture, which separates the computation and memory units and uses high-precision calculations  \cite{grollier2020neuromorphic}. Convolutional neural networks (CNN) are a class of artificial neural networks (ANNs) \cite{o2015introduction} which produces excellent performance in machine learning problems dealing with image data \cite{simonyan2014very}, computer vision \cite{khan2018guide,xie2017genetic}, and natural language processing \cite{albawi2017understanding}. \\
\indent A CNN has two stages: the first one is the feature extraction stage, and the second is the fully connected stage. Feature extraction is achieved by employing various layers, such as the convolutional layer, the activation function layer, and the pooling layer. These layers enable the CNN to exploit the spatial features of an image, introduce non-linearity \cite{sharma2017activation}, and suppress noise\cite{goodfellow2016deep}. \\
\indent The pooling layer is responsible for reducing the feature's size, which helps reduce the parameters and computation in the network. There are two types of pooling: max pooling and average pooling. Max pooling calculates the maximum value from the portion of the image feature, while the average pooling finds the average value of the part of the image \cite{boureau2010theoretical, o2015introduction}. The max-pooling also performs noise suppression by discarding noisy activations and is more often used in the network compared to the average pooling\cite{goodfellow2016deep}. \\
\indent The activation function introduces non-linearity to the network. Non-linearity enables the network to learn complex structures in the data and differentiates between outputs \cite{sharma2017activation}. Traditionally, sigmoid and tanh activation functions have been widely utilized. But, sigmoid and tanh functions saturate when the input is very high or low and are only sensitive to changes around their mid-points. After saturation, the network won't be able to learn well \cite{goodfellow2016deep,glorot2011deep}. The sigmoid and tanh functions also face the vanishing gradient problem, where the gradient information used to learn networks vanishes for deep networks. Without any helpful gradient information, deep networks won't be able to learn effectively \cite{goodfellow2016deep}. The vanishing gradient and saturation problems faced by tanh and sigmoid functions can be overcome by the rectified linear unit (ReLU) activation function\cite{goodfellow2016deep}.

\begin{figure*}[] 
\centering

  \begin{minipage}[c]{0.67\textwidth}
    \includegraphics[width=\textwidth]{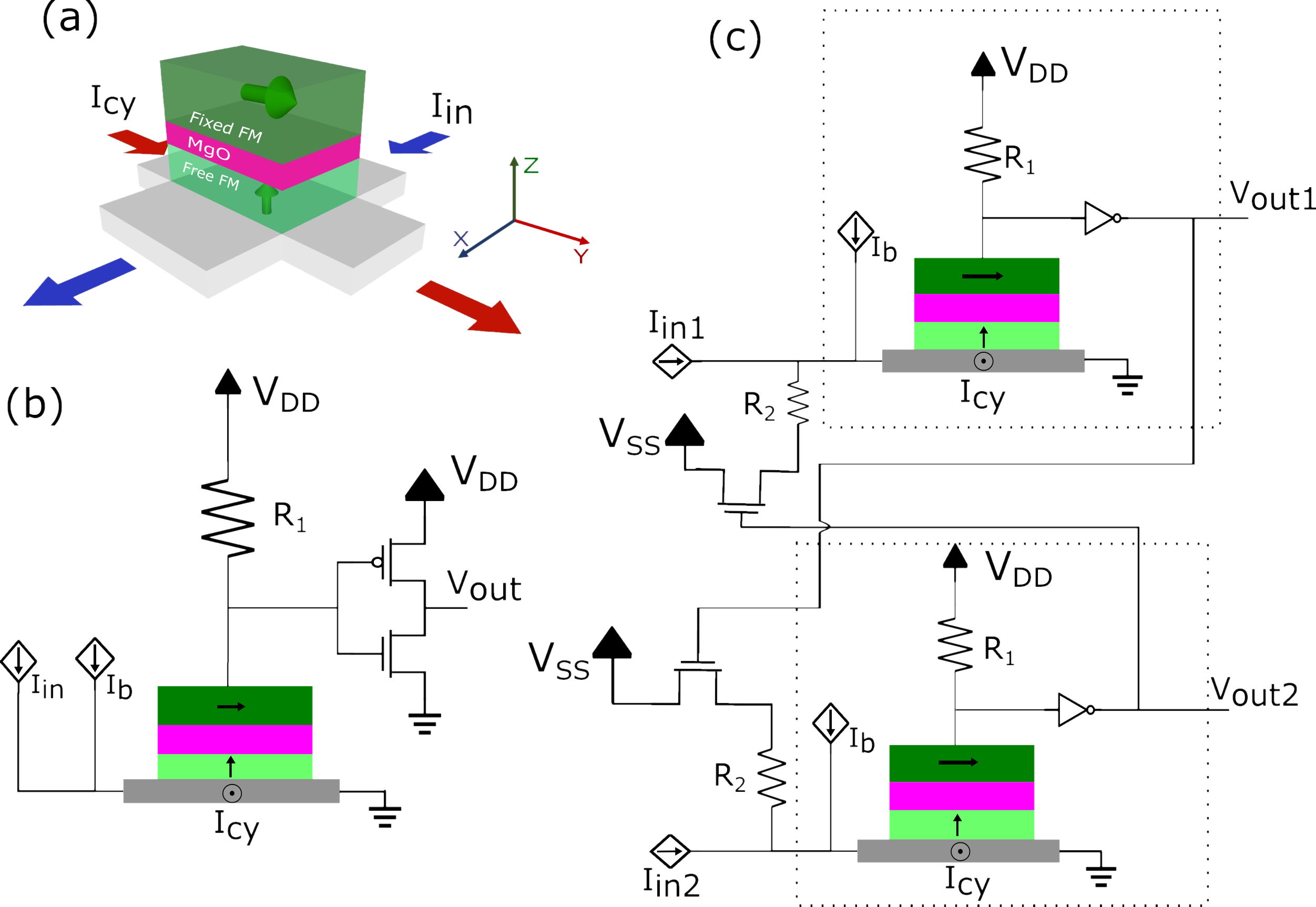}
  \end{minipage}\hfill
  \begin{minipage}[c]{0.3\textwidth}
   \caption{Design schematics. (a) the MTJ device on top of a SHE layer. Charge currents are injected along $\hat{x}$ and $\hat{y}$ directions, which produce a spin current along the $\hat{z}$-direction with spin-polarizations along $\hat{y}$ and -$\hat{x}$ directions respectively. (b) Circuit diagram for ReLU output. The MTJ and the resistor $R_1$ form a voltage divider whose output is fed to the CMOS inverter to produce the ReLU output. (c) Circuit diagram for local max-pooling function comprising two ReLU circuits. The inverted outputs of these ReLU circuits are fed as inputs to all other circuits, thus enabling the competition required for the max-pooling functionality.} 
    \label{fig:circuits}
  \end{minipage}
\end{figure*}
The hardware implementation of various layers of the feature extraction stage of a CNN has been explored by a few studies \cite{chang2019hardware, geng2020analog, priyanka2018cmos}, with the ultimate aim of developing neuromorphic computing \cite{schuman2017survey,furber2016large}. These works are based on CMOS circuit realizations of only the activation function \cite{chang2019hardware, geng2020analog, priyanka2018cmos} and lack the concatenation ability to perform simultaneous max-pooling. Furthermore, the in-memory computation using the existing CMOS technology is severely limited by area and energy requirements\cite{grollier2020neuromorphic, indiveri2011neuromorphic}. Spintronics, on the other hand, provides a wide range of devices and physical effects for in-memory computing that suits the hardware realization of neuromorphic computing and enables the paradigm of ``let the physics do the computing" \cite{parihar2017computing}. Although the spintronic implementation of activation functions is studied \cite{cai2019voltage,sengupta2015spin,cai2019sparse,raimondo2021reliability} but the implementation of max-pooling is still elusive, and moreover, some of these works require an external magnetic field to obtain the activation function. We propose a technologically relevant spintronic implementation of ReLU and max-pooling network using orthogonal spin current injection in MTJ device and show that unstable ferromagnets can be used for energy-efficient design employing an atom-to-circuit approach that weaves quantum transport modeling with CMOS circuit design.

Most of the works centered around hardware realization of neuromorphic computing are implementing a part of the network on hardware \cite{desai2022chip,borders2016analogue,sahu2022ferrimagnetic,singh2021learning,abbas2017memristor,hu2016dot,li2021hardware}  (like synapses based on domain wall, MTJ, nano-oscillators) and employ software-implementation (for activation function and pooling layers) to complete the network. The lack of spintronic hardware implementation of the activation function and pooling layers is a ramification of the absence of continuous change in the resistance of the MTJ device using spin current. Our work fills this critical gap by providing hardware for activation and max-pooling functions, thus enabling a fully-hardware implementation of neuromorphic networks.

\indent The paper is organized as follows. In Sec. \ref{design}, we describe the basics of the ReLU and the max-pooling functionalities. In Sec.\ref{simul}, we discuss in detail our developed hybrid simulation platform based on Keldysh non-equilibrium Green's function (NEGF) formalism coupled with the Landau-Lifshitz-Gilbert-Slonczewski (LLGS) equations for quantum transport description at the device level, which is then interfaced with the HSPICE circuit simulator to capture the spin and charge current interplay along with CMOS devices. In Sec.\ref{results} we present the performance of the ReLU and max-pooling circuit in terms of power consumption and noise analysis. We show that there exists a non-trivial power-error trade-off in the proposed ReLU-max-pooling network, which enables an energy-efficient network design based on unstable free ferromagnets with reliable outputs. We conclude in Sec. \ref{conclu}.

\section{Design}
\label{design}

\subsection{ReLU circuit}
The ReLU function emulation requires a device with linear characteristics. We propose a spin Hall effect (SHE)-driven MTJ device with orthogonal currents, as shown in Fig. \ref{fig:circuits}(a), to generate a linear and continuous rotation in magnetization of the free FM layer. In this work, we have used typical CoFeB-based FMs for the fixed and free layers with the equilibrium magnetizations along the $\hat{y}$-direction and the $\hat{z}$-directions, respectively. The fixed FM has in-plane magnetic anisotropy, and the free FM has a perpendicular magnetic anisotropy (PMA) \cite{sharma2021proposal} in order to achieve the respective equilibrium magnetization directions. The SHE layer generates two spin currents with orthogonal polarizations corresponding to the direction of input charge currents. These spin currents interact with the free FM layer to rotate the magnetization to the in-plane direction. The continuous rotation in magnetization can be translated to the electrical signal using the TMR of the MTJ device. In order to achieve the ReLU functionality, the SHE-driven MTJ is connected to a resistor to form a voltage divider, as shown in Fig. \ref{fig:circuits}(b). The voltage divider drives a CMOS inverter pair that operates in the linear region to produce the ReLU output. \\
\indent Continuous-linear rotation in the magnetization of the free FM layer can also be achieved by applying a magnetic field perpendicular to the uni-axial anisotropy field of the free FM \cite{sharma2016ultrasensitive}, as further elaborated in Appendix \ref{appendix:a}. But it increases the size and power consumption of the device \cite{apalkov2016magnetoresistive}. A recent work \cite{stephan2020spin} explored the ReLU functionality using Heusler alloy FMs and assuming only a field-like torque from the SHE layer, which may not give the desired linear rotation (see appendix: \ref{appendix:b}).

\begin{figure*}[]
	\centering
     \includegraphics[width=0.999\textwidth]{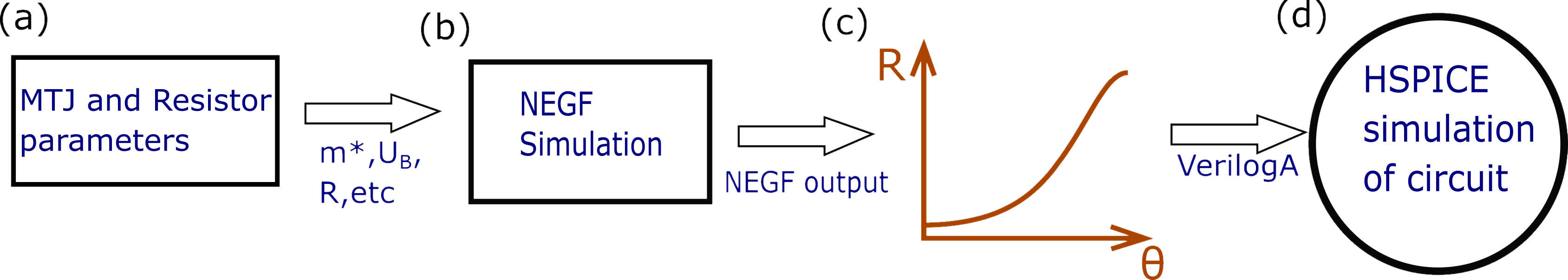}
	\quad
	\caption{Schematic overview of the simulation setup. (a) Various parameters of the MTJ and the resistor $R_1$ are given as input to the NEGF simulator. (b) The MTJ resistance is calculated using the NEGF framework, and the voltage across the MTJ is evaluated from the voltage divider self-consistently. (c) A schematic of the dependence of the resistance of the MTJ with the FM polarization angle is shown. (d) The NEGF results are incorporated into the HSPICE using VerilogA, and the entire circuit is simulated, including the LLGS equation.}
	\label{fig:flow}
\end{figure*}

 %In order to achieve the ReLU functionality, the SHE-driven MTJ is connected to a resistor to form a voltage divider, as shown in Fig. \ref{fig:circuits}(b). The voltage divider drives a CMOS inverter pair that operates in the linear region to produce the ReLU output. The orthogonal injection of spin currents in the free layer which generates a  continuous rotation of magnetization of the free FM layer causes the relative angle between the magnetizations of the fixed and free layer to vary between 180\degree (anti-parallel configuration) to 0\degree (parallel configuration) with injected currents (see Fig \ref{fig:linear}). The TMR translates the change in the angle to a reduction in the resistance of the MTJ, as shown in Fig. \ref{fig:flow}(c). The fall in the MTJ resistance decreases the voltage at the input of the CMOS inverter. The CMOS inverter biased in the linear region amplifies and inverts the voltage signal, resulting output of the inverter as the ReLU output shown in Fig. \ref{fig:relu}.
 
\subsection{Max-Pooling}
\label{max_section}

 Max-pooling forms a crucial layer for the CNN, and we discuss in this section that the proposed ReLU circuit can be appropriately concatenated to perform the simultaneous local max-pooling function. The max-pooling function calculates the maximum of the inputs presented \cite{o2015introduction}. We introduce a competition among the ReLU circuits to achieve max-pooling with only one winner. We present a strategy that makes the current input to all the ReLU circuits (except one) less than zero to have one winner. We will achieve this by drawing current from the input of the ReLU circuits. The magnitude of the current drawn depends on the output of other ReLU circuits, thus enabling competition. At the end of the operation, there will be only one ReLU circuit with non-zero output. This ReLU output injects a negative current to all other ReLU devices so that their effective input remains less than zero. \\
\indent Our design enables the competition through an NMOS transistor connected to a resistor. The gate terminal of the transistor connects to the output of a different ReLU circuit. Thus, the current drawn from one ReLU circuit depends on the output voltage of another ReLU circuit. Figure \ref{fig:circuits}(c) shows the schematic of the ReLU+max-pooling circuit with two inputs. This circuit design calculates both the ReLU and max-pooling functions simultaneously. This strategy makes all outputs zero other than the one corresponding to the maximum input. We can sum all the outputs and get the correct result without wondering which device has the maximum input. \\
\begin{figure}[b]
	\centering
    \includegraphics[height=0.25\textwidth,width=0.4\textwidth]{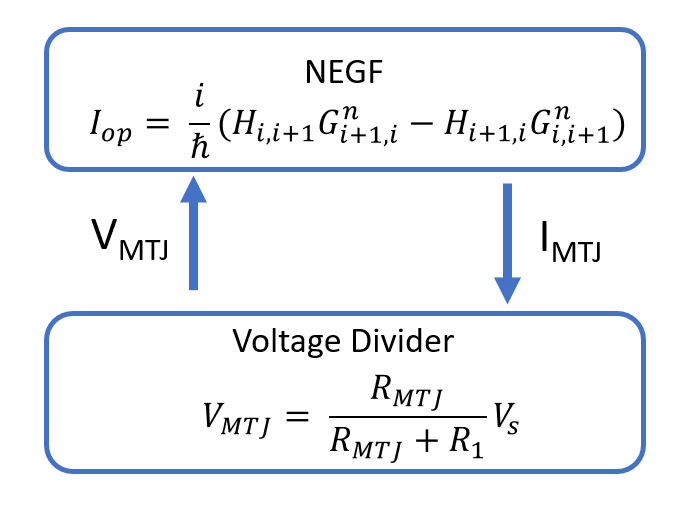}
	\caption{Schematic of the device simulation section. The NEGF quantum transport formalism is self-consistently coupled with the voltage divider circuit to calculate the MTJ resistance as a function of the FM polarization angle. Here $H$ is the device Hamiltonian and $G^n$ is the electron correlation matrix, see appendix for more details.}
	\label{fig:negf_volt_divi}
\end{figure}

\indent Crossbar arrays \cite{jo2010nanoscale,fan2014design} can be used to calculate the convolution function, and the output of the convolution layer is given as input to the proposed simultaneous ReLU-max-pooling circuit to complete the feature extraction stage in the CNN as shown in Fig. \ref{fig:full_cnn}.
Non-volatile memory devices such as domain-wall \cite{siddiqui2019magnetic,grollier2020neuromorphic} and skyrmion-based \cite{huang2017magnetic,grollier2020neuromorphic} MTJs can be used as synapses in the crossbar arrays. The input resistance of the proposed network can be tuned (by changing the dimensions of the SHE layer) as per the loading requirement of the crossbar array. The proposed ReLU circuit can be directly connected to the crossbar array without the max-pooling portion to complete the McCulloch-Pitts neuron\cite{goertzel2013structure} that can be used in hidden layers of feed-forward networks \cite{sandberg2001nonlinear,svozil1997introduction} and reservoir computing \cite{nakajima2021reservoir}.

\label{simul}
\begin{table}
	\centering
\caption{ Simulation Parameters.}
	\label{table1}
	\begin{tabular}{|l| >{}m{4cm} | >{}m{3cm} |}
		\hline
\textbf{Symbol} & \textbf{Quantity} & \textbf{Value}\\ %[10pt]
\hline
 $M_s$ & saturation magnetization & 1150 emu/$cm^3$ \\ 
 \hline
$H_k$ & anisotropy field & 330 - 3300 Oe\cite{gajek2012spin} \\ 
 \hline  
 V & volume of ferromagnet & 1000 $nm^3$\\ 
 \hline
 $\Delta$ & thermal stability factor & 4.58 - 45 \\ 
 \hline
 $\alpha$ & Gilbert damping & 0.01\\ 
   \hline
L & length of ferromagnet & 40 nm\\
   \hline
W & width of ferromagnet & 25 nm\\
\hline
$C_{MTJ}$ & MTJ capacitance & 2.21 fF \\
   \hline
 $I_{cy}$ & input current to SHE layer & 35 - 340 $\mu A$\\
 \hline
 $\theta$ & spin-hall angle & 0.3 rad \cite{pai2012spin}\\
  \hline
 $t_{HM}$ & thickness of heavy metal & 5 nm\\
% \hline
% $R_{HM}$ & heavy metal input resistance & 21 $\Omega$\\
 
  \hline
$R_1$ & {reference resistor} & 698.63 k$\Omega$\\
\hline
$R_2$ & {resistance in NMOS current source} &1 - 7 k$\Omega$\\
\hline
$I_b$ & biasing current & 17 - 160 $\mu A$\\
  \hline

$V_{DD}, V_{SS}$ & voltage sources & 0.5 V, -0.5 V\\
\hline
$C_g$ & CMOS inverter input capacitance& 0.175 fF \\
\hline
$C_o$ & CMOS inverter output capacitance& 0.305 fF \\
  \hline
$\Delta  t$ & simulation time step & 0.5 pS\\
\hline
$\hbar$ & reduced Plank's constant & $1.055\times 10^{-34}$ J s\\
\hline
$k_B$ & Boltzmann constant & $ 1.38\times10^{-16} erg K^{-1}$\\
\hline
T & temperature & 300 K\\
\hline
	\end{tabular}
		
\end{table}

\section{Simulation Methods}
\label{simul}

We show in Fig. \ref{fig:flow} the schematic overview of the developed hybrid NEGF-CMOS simulation framework. The MTJ parameters, such as the effective mass of electrons in the insulator and the FM, the barrier height of the insulator, thickness, area, and the resistance value of $R_1$ are presented to the NEGF simulator as shown in Fig. \ref{fig:flow}(a).
The resistance of the MTJ depends on the MTJ angle and the voltage across MTJ\cite{sankey2008measurement,wang2009bias}, this is shown in Fig. \ref{fig:vmtj_resist}. This effect is substantial, and using a simple conductance equation to describe the MTJ ignores this effect. NEGF, on the other hand, includes these effects via the atomistic level simulation whose results agree with the experimental data\cite{sankey2008measurement,kubota2008quantitative,miao2006inelastic}. In circuits employing MTJ, there is usually a change in MTJ voltage. The combination of these effects stresses the necessity for using NEGF self-coupled with the circuit simulation to capture the resistance dependence on the voltage. While simulating CMOS devices such as inverters, HSPICE is very useful, and integrating NEGF with HSPICE provides a pathway to design circuits with spintronic and CMOS devices. This integration also allows access to HSPICE's analysis tools, such as noise and frequency analysis. This coupling of NEGF and HSPICE is further helpful in simulating more extensive networks.
The NEGF simulator calculates the resistance of the MTJ for a given MTJ orientation angle and voltage across it. The voltage across the MTJ depends on the voltage divider formed by the MTJ and resistor $R_1$, so the NEGF and voltage divider are simulated self consistently as shown in Fig. \ref{fig:negf_volt_divi}.

\indent This result is coupled to the HSPICE simulator using the VerilogA code. The HSPICE simulates the entire circuit, including the LLGS equation \cite{panagopoulos2013physics,sun2000spin}, and the HSPICE also simulates the approximation of the CMOS inverter pair based on a 16nm node of the predictive technology model (PTM) \cite{Predicti46:online}. The circuit parameters used in the simulation are given in Tab.\ref{table1}.
We have employed The Keldysh NEGF formalism\cite{datta1997electronic,sharma2017resonant} coupled with LLGS equations\cite{slonczewski1996current,brataas2012current,sharma2018role} to describe the magnetization dynamics of the free FM(see appendix: \ref{appendix:siml} for more details on the simulation methods).

\section{Results}
\label{results}

\begin{figure}[]
	\centering
 
    \subfloat[]{\includegraphics[width=0.49\linewidth]{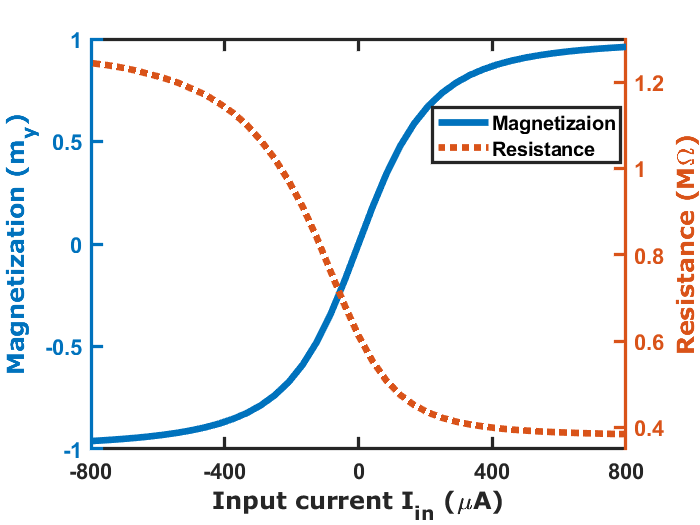}\label{fig:linear}}
	\subfloat[]{\includegraphics[width=0.49\linewidth]{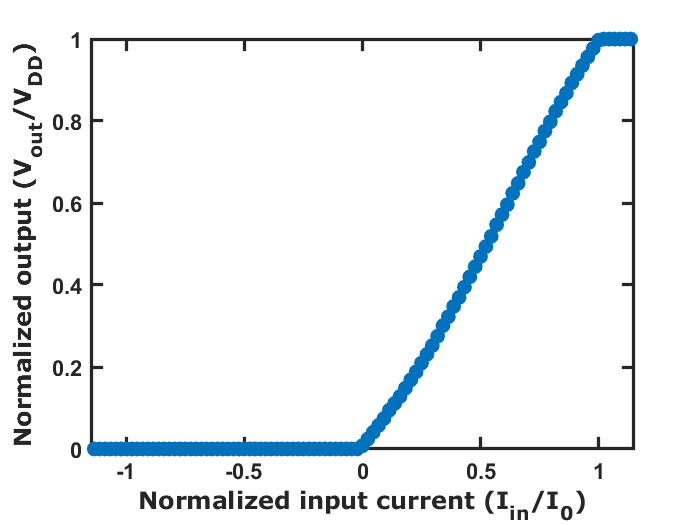}\label{fig:relu}}
	\caption{Magnetization dynamics and the ReLU realization. (a) The magnetization($m_y$) of the free FM layer and the resistance of the MTJ-device with the current $I_{in}$ along $\hat{x}$-direction, while a constant critical orthogonal current of 340 $\mu A$ is applied along $\hat{y}$-direction (see Fig.~\ref{fig:circuits}(a)). (b) The normalized output voltage of the ReLU circuit (see Fig.~\ref{fig:circuits}(b)) with normalized input current with $I_0 = 220 \mu A$. The thermal stability factor for the given ReLU is 45.}
	\label{meh}
\end{figure}

We show in Fig. \ref{fig:linear} the continuous rotation of the magnetization of the free FM layer with the input current $I_{in}$, the stability of the magnetization is discussed in appendix \ref{appendix:mag_stability}. The continuous rotation of the magnetization with $I_{in}$ is achieved when the orthogonal current $I_{cy}$ magnitude is larger than a critical value (see appendix \ref{appendix:c}). The TMR effect translates the continuous change in the magnetization to a continuous resistance change of the MTJ device, as shown in Fig. \ref{fig:linear}. The linear region of the MTJ-resistance around the zero $I_{in}$ has been utilized to obtain the ReLU output as shown in Fig.~\ref{fig:relu} by injecting an appropriate bias current $I_b$. Various parameters of the ReLU circuit design are given in Tab. \ref{table1}. The proposed circuit shown in Fig.~\ref{fig:circuits}(b) emulates the ReLU function closely for normalized inputs of smaller than unity.

\begin{figure}[]
	\centering
	\subfloat[]{\includegraphics[width=0.49\linewidth]{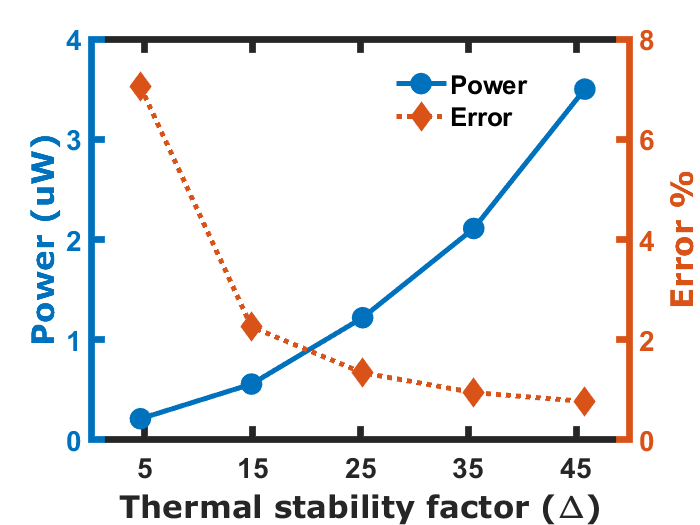}\label{fig:relu_power_error}}
	\subfloat[]{\includegraphics[width=0.49\linewidth]{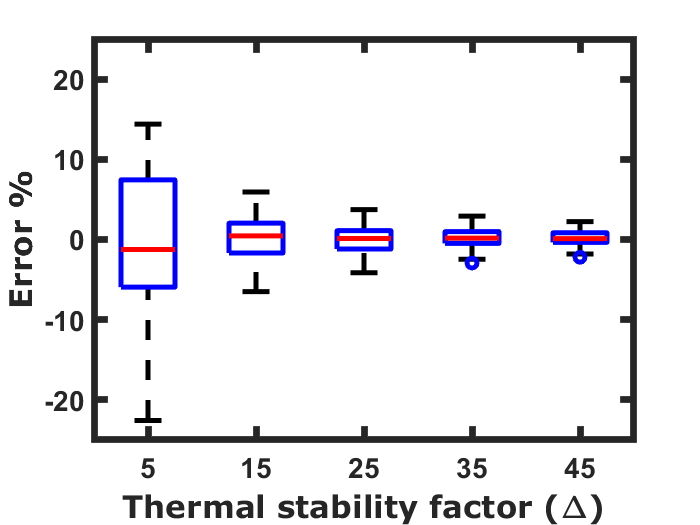}\label{fig:relu_error_box}}
	\caption{ Trade-off between power consumption and error percentage. (a) Static power consumption of the ReLU circuit over the entire input range and the average of absolute error percentage with normalized input ($I_{in}/I_0$) of 0.5. (b) Boxplot of the error percentage of the ReLU circuit under a thermal noise. The error percentage increases with a decrease in the thermal stability factor.}
	\label{meh}
\end{figure}

\indent We evaluate the performance of the ReLU circuit against the thermal stability factor ($\Delta = \frac{H_k M_s V}{2 k_B T}$) of the free FM layer of the MTJ device. The $\Delta$ factor of the free-FM not only captures the stability of the magnetization direction against thermal noise but also the extent to which the spin current changes the magnetization direction. The critical spin current ($I_{sc}=\frac{4e\alpha k_B T\Delta}{\hbar}$) for magnetization switching is proportional to $\Delta$ \cite{sharma2021proposal}. We vary the $\Delta$ of the free FM layer by changing the anisotropy field, but the same can also be changed by the MTJ device dimensions. The critical orthogonal current ($I_{cy}$), bias current ($I_b$), and the input current($I_{in}$) required for a continuous rotation of magnetization decreases with a reduction in $\Delta$. Thus, the mean static power consumption of the ReLU circuit also decreases with a reduction in $\Delta$ as shown in Fig. \ref{fig:relu_power_error}. We also observe that the $\Delta$ reduction does not affect the output settling time ($\approx$4ns) as various input currents decrease proportionally with $\Delta$. \\
\indent We show in Fig. \ref{fig:relu_power_error} the reduction in the output error of the ReLU circuit with a corresponding increase in $\Delta$. The thermal noise ($ \langle H_{th}^2 \rangle = \frac{2\alpha k_B T}{\gamma M_s V}$) stays constant as $\Delta$ is increased and the weight of the thermal noise decreases with $H_{eff}$ as $H_k$ is increased, since $H_{eff}$ also includes the contribution from the anisotropy field $H_k$. 
The error in the output is estimated by performing fifty cycles of Monte Carlo  simulations \cite{hammersley2013monte} with a normalized input current ($I_{in}/I_0$) of value $0.5$. The decrease in power consumption and the increase in the error due to the reduction in $\Delta$ presents an opportunity to optimize the circuit to consume less power while keeping the error in an acceptable range. It can be inferred from Fig. \ref{fig:relu_power_error} that the ReLU circuit has an optimal performance of power($0.5-1.2 mW$) and error($<2.2\%$) with $\Delta$ in the range of $15-25$.

\indent Figure~ \ref{fig:relu_error_box} shows the boxplot of the error (\%) with the $\Delta$. The boxplot \cite{hubert2008adjusted} gives profound insights into the error statistics and uses the following five parameters to summarize the data error. The medians (the red lines) converging to zero indicate that the average thermal noise is zero. The reduction in its spread is associated with the variance of the thermal noise since its effect on free-FM depends on $\Delta$.

\begin{figure}[]
	\centering
	\subfloat[]{\includegraphics[width=0.49\linewidth]{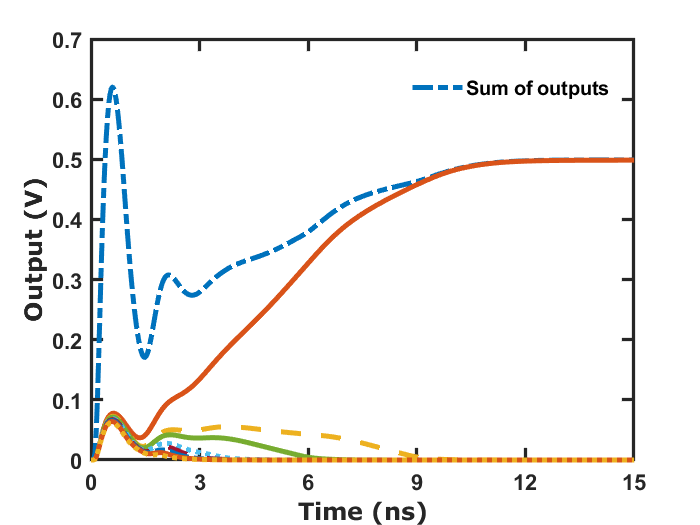}\label{fig:max_transient}}
	\subfloat[]{\includegraphics[width=0.49\linewidth]{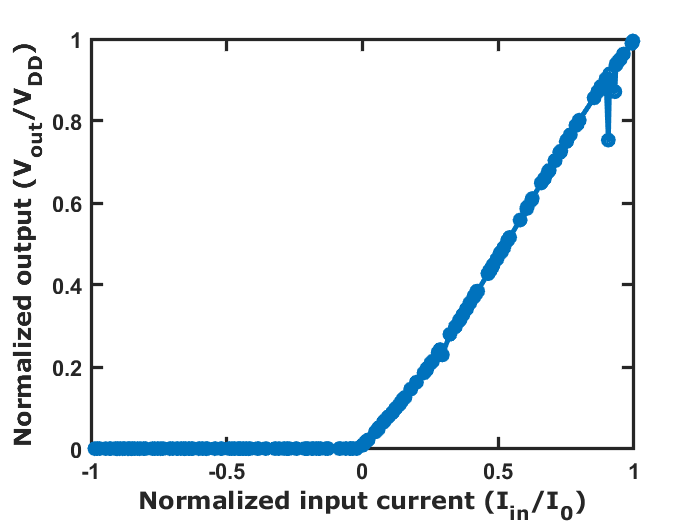}\label{fig:max}}
	\caption{Max-pooling circuit transient response. (a) The transient response of the ReLU-max-pooling network with nine inputs. The inputs are chosen to demonstrate the worst-case scenario. (b) The output of the ReLU-max-pooling network with nine inputs, the x-axis is the maximum input current out of all the nine inputs normalized to $I_0 = 220 \mu A$.}
	\label{meh}
\end{figure}

\indent We show in Fig. \ref{fig:max_transient} the transient response of the ReLU-max-pooling circuit with 9-inputs corresponding to the typically used $3 \times 3$ pooling layer for the worst-case scenario. In the worst-case scenario, all the inputs are close to the maximum possible value and the observed settling time of the output is $9 ns$. Initially, all the nine outputs rise to reach the output dictated by the corresponding inputs. Then quickly, the competition starts, and all the ReLU circuits inject negative currents into each other through the NMOS transistors causing the outputs to decrease. As the competition progresses, one of the outputs (that has the maximum input) becomes the winner and reaches its steady-state value corresponding to its input while keeping all the other outputs at zero. It can also be inferred from Fig \ref{fig:max_transient} that the sum of all the outputs is equal to the maximum output in the steady-state, and it performs the max-pooling without any knowledge about which of the 9-ReLU devices is producing it.

\begin{figure}[!t]
	\centering
	\subfloat[]{\includegraphics[width=0.49\linewidth]{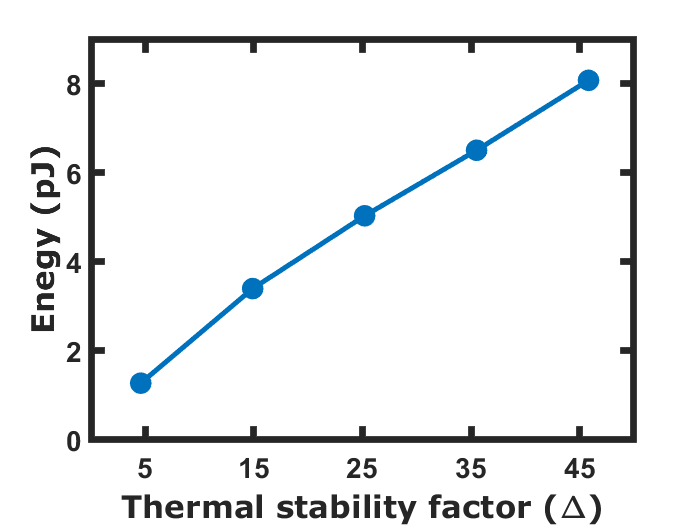}\label{fig:max_energy}}
	\subfloat[]{\includegraphics[width=0.49\linewidth]{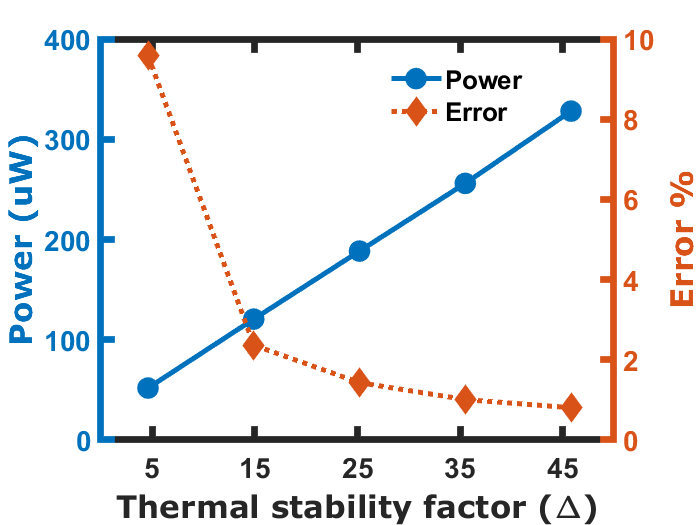}\label{fig:max_power_error}}\\
	\subfloat[]{\includegraphics[width=0.55\linewidth]{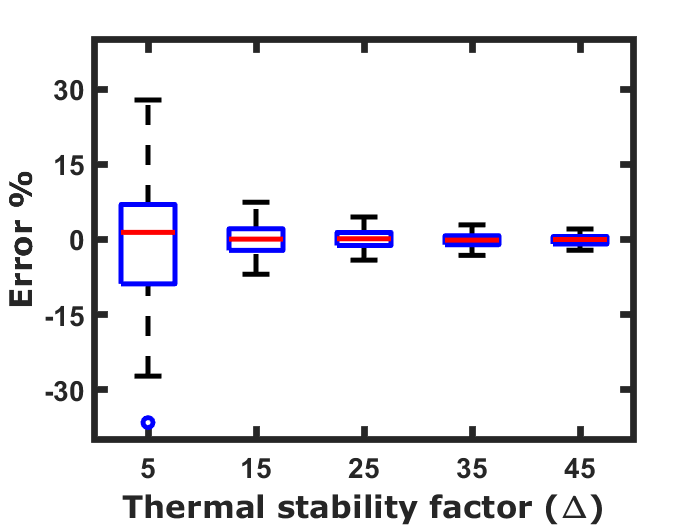}\label{fig:max_error_box}}

	\caption{Energy requirements. (a) Worst-case energy requirement for the output to settle for a 9-input ReLU-max-pooling network. (b) Static power consumption of a 9-input ReLU-max-pooling circuit averaged over the entire range of inputs and the average of absolute error percentage of the network for a normalized input of 0.5. (c) Boxplot of the error percentage of a 9-input network (see Fig. \ref{fig:circuits}(c)) under thermal noise.}
	\label{maxxx}
\end{figure}

\indent Figure \ref{fig:max} shows the ReLU-max-pooling network results for $200$ Monte Carlo simulations. The value of the input currents is varied using the Monte Carlo method. The inputs are selected such that the maximum input is varied over the entire input range. The results from Fig. \ref{fig:max} closely resemble the ReLU function while simultaneously calculating the max pooling. \\
\indent We show in Fig. \ref{fig:max_energy} the ReLU-max-pooling network's energy consumption to reach the steady-state in the worst-case scenario. The worst-case energy consumption decreases as $\Delta$ decreases. This decrease in energy is due to the reduction of energy requirement for each ReLU circuit and the reduction in the competition circuit energy that is caused by a decrease in the NMOS device dimensions and resistance $R_2$ due to the decrease in $I_{in}$ required.

\indent Figure~\ref{fig:max_power_error} shows the static power consumption and percentage error of the network. The static power consumption decreases, and the percentage error increases with a reduction in $\Delta$. This behavior shares the same rationale as that of a single ReLU device discussed earlier. As $\Delta$ increases, there is a linear increase in power consumption and a sharp fall in error ($\%$). It opens up a possibility for the optimization of the power consumption ($120\mu W$) with an acceptable percentage error ($2.3\%$) at $\Delta\approx15$. \\
\indent The above discussion suggests that a room temperature unstable free FM ($\Delta<40)$ based MTJ can be utilized to design an energy-efficient ReLU-max-pooling network. We show in Fig. \ref{fig:max_error_box} the boxplot of the error percentage of the 9-input network. Here the medians converge to zero, indicating that the average thermal noise is zero. As $\Delta$ increases, the effect of thermal noise on the free-FM decreases, causing a decline in the spread of the boxplot.

\section{Conclusion}
\label{conclu}
In this paper, we proposed a circuit design for calculating crucial neuromorphic functions (ReLU and max-pooling) based on a continuous rotation of magnetization of the free FM of an MTJ through orthogonal spin current injection. Using our developed simulation platform based on NEGF and LLGS equations coupled with the HSPICE circuit simulator, we showed the optimal range of performance (power and error) against the thermal stability factor of the free FM of the MTJ device. We demonstrated that the designed circuit is robust against thermal noise, while consuming $0.56\mu W$ of power for ReLU functionality and requires $3.4 pJ$ of energy for 9-input ReLU-max-pooling computation in the worst-case scenario, and consumes 120 $\mu W$ of static power in a typical scenario with a percentage error of 2.3\% at $\Delta=15$. Our work also opens the possibility of using an unstable FM ($\Delta=15$) for an energy-efficient ReLU-max-pooling circuit with reliable output.

\section*{Acknowledgements}
The author BM acknowledges the Visvesvaraya Ph.D Scheme of the Ministry of Electronics and Information Technology (MEITY), Government of India, implemented by Digital India Corporation (formerly Media Lab Asia), and also acknowledges the support by the Science and Engineering Research Board (SERB), Government of India, Grant No. STR/2019/000030 and Grant No. MTR/2021/000388. We also acknowledge the Ministry of Human Resource Development (MHRD), Government of India, Grant No. STARS/APR2019/NS/226/FS under the STARS scheme. The author AS acknowledges the support of ISIRD phase-1 project of IIT Ropar.

\section*{APPENDIX}
%\appendix
%\counterwithin{figure}{section}
\renewcommand{\thefigure}{A\arabic{figure}}
\setcounter{figure}{0}

\renewcommand*{\thesection}{A\arabic{section}}
\setcounter{section}{0} 

\section{CNN architecture}
\begin{figure*}[]
	\centering
     \includegraphics[width=0.999\textwidth]{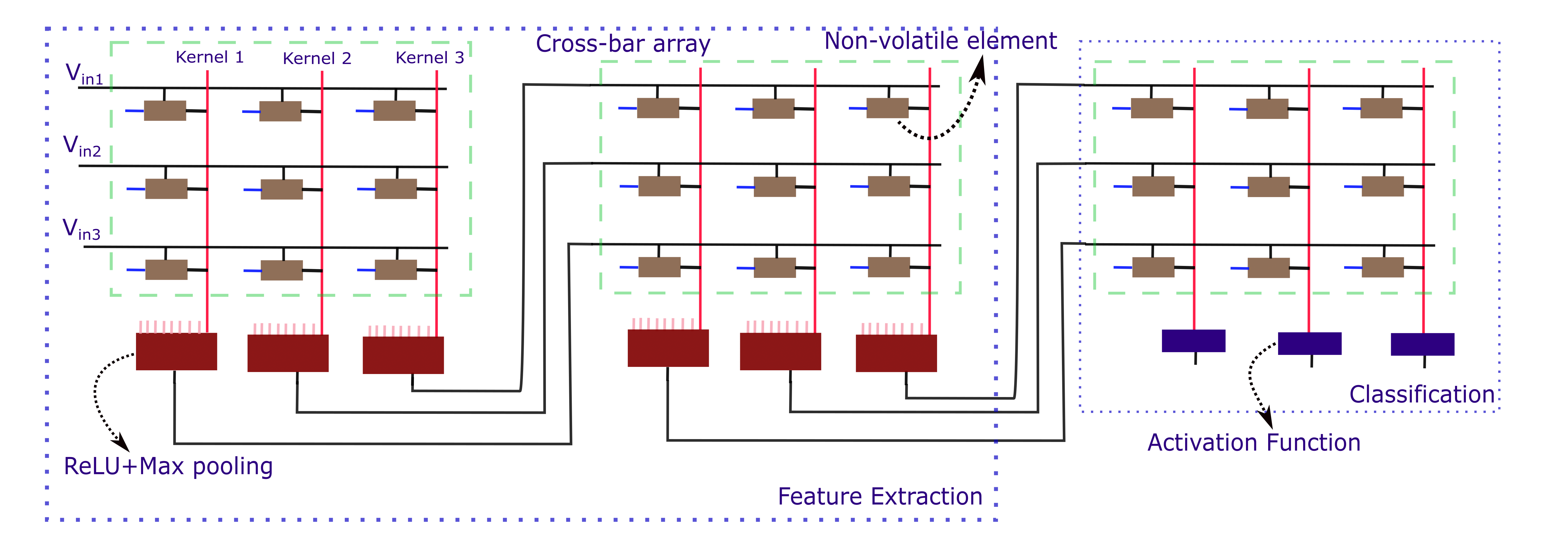}
	\quad
	\caption{An example of CNN architecture using crossbar arrays and ReLU-max pooling circuits. The combination of the crossbar array and ReLU-max pooling circuit is repeated multiple times (depending on the complexity of the problem) to complete the feature extraction stage of CNN, whose result is then given to the classification stage to complete the CNN network.}
	\label{fig:full_cnn}
\end{figure*}

An example of CNN architecture using crossbar arrays and ReLU-max pooling circuits is given in Fig. \ref{fig:full_cnn}. Here The combination of the crossbar array and ReLU-max pooling circuit is repeated multiple times (depending on the complexity of the problem) to complete the feature extraction stage of CNN, whose result is then given to the classification stage to complete the CNN network.

\section{MTJ resistance dependence on bias voltage}
\begin{figure}[]
	\centering
	\subfloat[]{\includegraphics[width=0.49\linewidth]{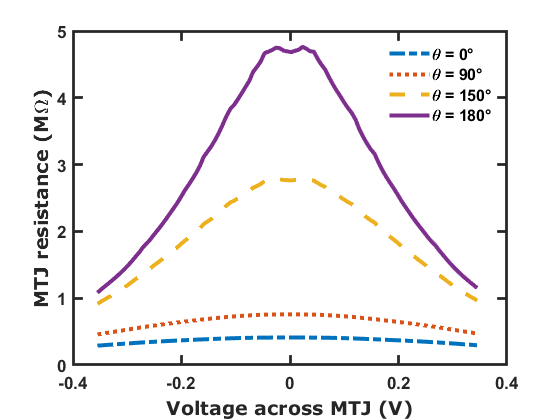}\label{fig:vmtj_resist}}
	\subfloat[]{\includegraphics[width=0.49\linewidth]{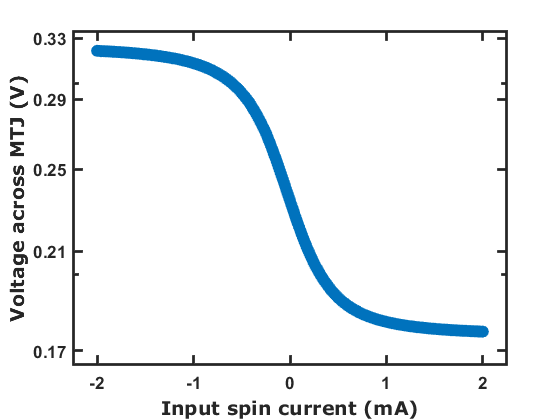}\label{fig:Ispin_vmtj}}
	\caption{(a) Change in resistance of the MTJ due to variation in voltage for different MTJ angles ($\theta$).  (b) The voltage across the MTJ with varying input spin current. }
	\label{meh}
\end{figure}

The resistance of the MTJ depends on the MTJ angle and the voltage across MTJ as shown in Fig. \ref{fig:vmtj_resist}. The voltage change with respect to the input current in the proposed ReLU circuit is shown in Fig.\ref{fig:Ispin_vmtj}.

\section{Simulation Method}

\subsection{Quantum Transport}
\label{negf}

The Keldysh NEGF technique is a versatile formalism used to describe quantum transport across a variety of set ups featuring charge, spin, heat flow \cite{datta1997electronic,datta2012modeling,datta2018lessons,sharma2017resonant} and also can be used to describe transport across hybrid quantum systems featuring the interplay of transport with the physics of dissipation and incoherent processes involving phonons, photons and also Cooper pairs \cite{Singha,Urs,Yeyati,San_Jose_2013,Praveen,Duse_2021}.
We adapt this formalism \cite{sharma2017resonant} to simulate the MTJ structure with MgO sandwiched between free and fixed FM layers. For a tight-binding Hamiltonian of the CoFeB, the parameters used are exchange splitting $\delta = 2.15 eV$, Fermi energy $E_f = 2.25 eV$, effective mass of MgO barrier is $m_{OX} = 0.18 m_e$ and that of FM contact is $m_{FM} = 0.8 m_e$ , where $m_e$ is the free electron mass. The barrier height of the CoFeB-MgO interface is $U_B = 0.76 eV$ above the Fermi energy \cite{datta2011voltage, kubota2008quantitative}, and we have adjusted the width of the MgO layer to 3nm so that spin-transfer torque has no impact on the magnetization dynamics of the free FM layer.

We start by describing the Green's function matrix $[G(E)]$ 

\begin{equation}
    [G(E)] = [EI-H-\Sigma]^{-1}
\end{equation}

\begin{equation}
    [\Sigma] = [\Sigma_T]+[\Sigma_B]
\end{equation}

Where [H] is the device Hamiltonian, $[H]=[H_0]+[U]$, comprising device tight-binding matrix $[H_0]$ and the Coulomb charging matrix $[U]$, and [I] is the identity matrix, $E$ is the energy variable. The self-energy matrix $[\Sigma]$ is the sum of the top $[\Sigma_T]$ and bottom $[\Sigma_B]$ FM layer's self-energy matrices. The charging matrix $[U]$ is calculated self-consistently using Poisson's equation.

The electron correlation matrix $[G^n(E)]$ and the in-scattering function $[\Sigma^{in}(E)]$ are given by

\begin{equation}
    [G^n]= \int dE [G(E)] [\Sigma^{in}(E)][G(E)]^\dagger
\end{equation}

\begin{equation}
    [\Sigma^{in}(E)] = [\Gamma_T(E)]f_T(E) + [\Gamma_B(E)]f_B(E)
\end{equation}

where, $[\Gamma_T (E)] = i([\Sigma_{T} (E)]-[\Sigma_T (E)]^\dagger)$ and $[\Gamma_B (E)] = i([\Sigma_B (E)]-[\Sigma_B (E)]^\dagger)$ represent spin-dependent broadening matrices of the top and bottom contacts.

The quantum transport segment \cite{datta1997electronic,sharma2017resonant} culminates with the calculation of currents via the current operator that describes the current flow between sites $i$ and $i+1$ is given by 
\begin{equation}
    I_{op} = \frac{i}{\hbar}(H_{i,i+1}G^n_{i+1,i} - H_{i+1,i}G^n_{i,i+1})
\end{equation}
This current operator $\hat{I}_{op}$ is $2 \times 2$ matrix in the spin space of a lattice point, using which the charge current can be evaluated as
\begin{equation}
    I= q \int Real [Trace(\hat{I}_{op})] dE,
\end{equation}
where $q$ is the quantum of electronic charge. We also consider the effect of MTJ capacitance since it dominates the CMOS inverter pair capacitance.

\subsection{Magnetization dynamics}
 We utilize the LLGS equation \cite{slonczewski1996current,brataas2012current} to describe the magnetization dynamics of the free FM. The LLGS equation is given by
\begin{multline}
  (\frac{1+\alpha^2}{\gamma H_k})\frac{d \hat{m}}{dt} = -\hat{m} \times \Vec{h}_{eff} - \alpha \hat{m} \times \hat{m} \times \Vec{h}_{eff} \\ - \hat{m} \times \hat{m} \times \Vec{i}_s + \alpha \hat{m} \times \Vec{i}_s,
\end{multline}
where $\hat{m}$ is the unit vector along the direction of magnetization of the free magnet, $\gamma$ is the gyromagnetic ratio, $\alpha$ is the Gilbert damping parameter, $\Vec{h}_{eff} = \frac{\Vec{H}_{eff}}{H_k}$ is the reduced effective field and $\Vec{i}_s= \frac{\hbar\Vec{I}_{s}}{2qM_sVH_k}$ is the normalized spin current. The term $\vec{H}_{eff}$ includes the contribution of the anisotropy field ($H_k$), the applied magnetic field ($H_{app}$) and the thermal noise ($H_{th}$). The thermal noise is given by $ \langle H_{th}^2 \rangle = \frac{2\alpha k_B T}{\gamma M_s V}$ and $\langle \rangle$ represents the ensemble average \cite{sun2004spin}.
\indent The spin current from the SHE layer is given by \cite{hirsch1999spin, takahashi2008spin}
\begin{equation}
    I_s = \theta \frac{l_{FM}}{t_{HM}} I_c \times \sigma
\end{equation}
where $I_s$ is the magnitude of the spin current, $\theta$ is the spin Hall angle, $l_{FM}$ is the length of the FM, $t_{HM}$ is the thickness of the SHE layer, $I_c$ is the charge current, and $\sigma$ is the polarization of the spin current.

\section{Continuous linear rotation in the magnetization}
In this section, we explore various possible routes for the continuous rotation of magnetization. We also examine the feasibility of each circuit design.

\subsection{Static magnetic field driven MTJ}
\label{appendix:a}
Here, we show that linear resistance can be generated from an MTJ by applying a static magnetic field orthogonal to the uni-axial anisotropy direction of the free FM.

\begin{figure}[]
	\centering
	\subfloat[]{\includegraphics[width=0.49\linewidth]{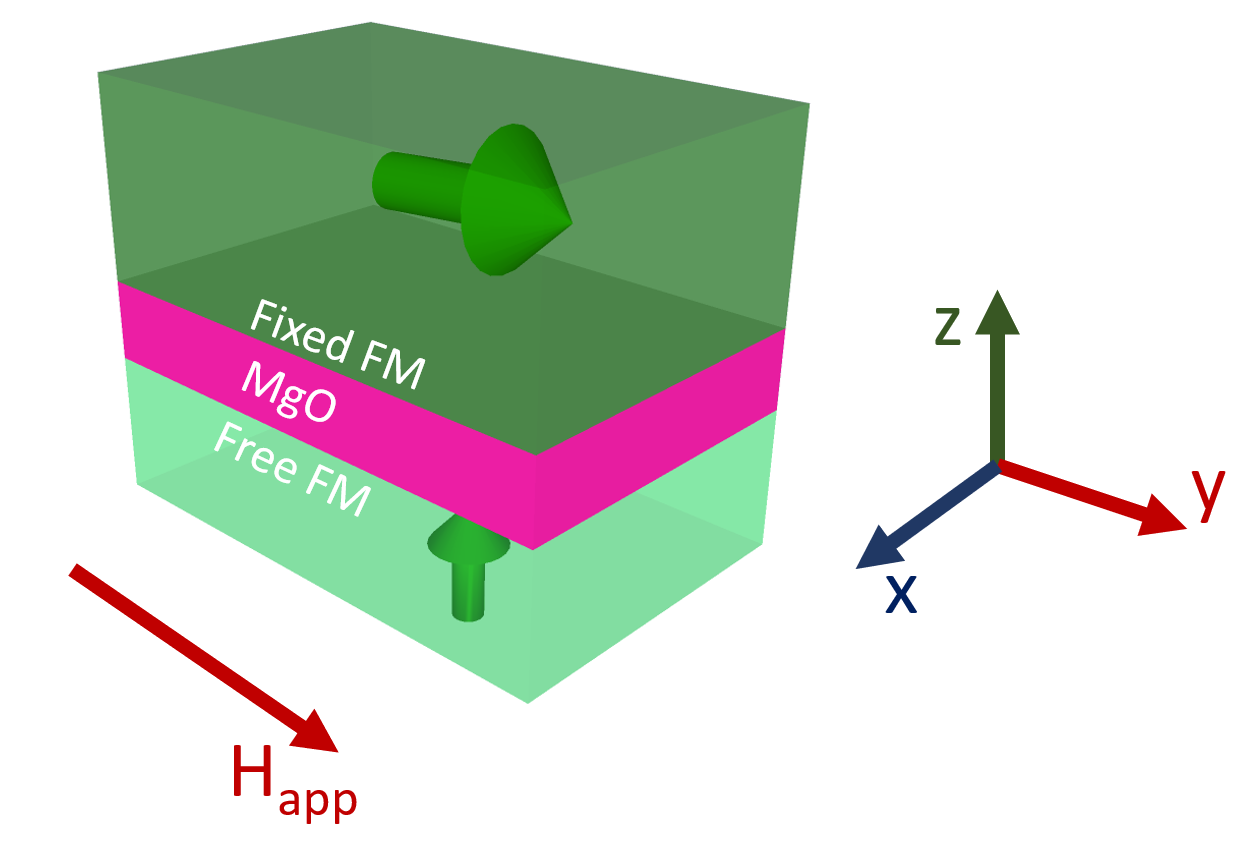}\label{fig:rweraer}}
	\subfloat[]{\includegraphics[width=0.49\linewidth]{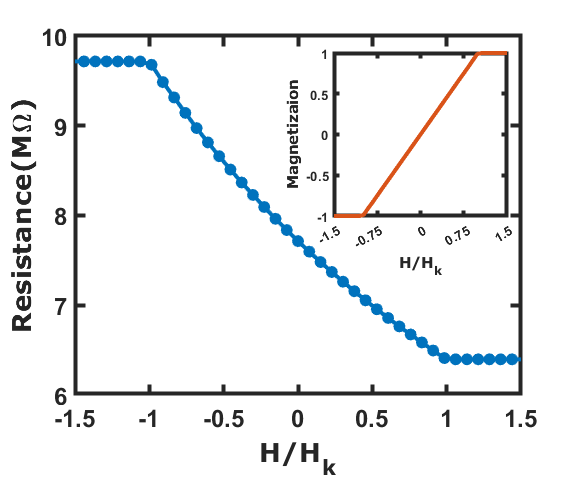}\label{fig:erwerfaser}}
	\caption{Device schematic. (a) Schematic of an MTJ device. The PMA magnet is the free FM layer, and the in-plane magnet is the fixed FM layer. The magnetic field is applied along the $\hat{y}$-direction to get a linear rotation in the magnetization of the free layer. (b) Variation of the MTJ resistance with the normalized magnetic field ($H/H_k$) with $V_{MTJ}=0.33 V$. Inset shows the variation of the magnetization ($m_y$) of the free FM layer with an applied magnetic field.}
	\label{mag_field}
\end{figure}

Fig. \ref{mag_field}(a) shows the schematic of MTJ with PMA FM as free layer and in-plane magnet as fixed FM layer. To generate linear rotation in magnetization, magnetic field is applied in the y-direction. Figure \ref{mag_field}(b) shows the simulation results of the MTJ. LLGS and NEGF formalism are employed in the simulation of the device setup.
This method of using magnetic field is not recommended because it increases the size and power consumption of the device \cite{apalkov2016magnetoresistive}.

\subsection{Field-like spin torque driven MTJ}
\label{appendix:b}
The LLGS equation describes the magnetization dynamics of the free FM. Equation \ref{eq:llg_main} shows the LLGS equation.

\begin{multline} \label{eq:llg_main}
  (\frac{1+\alpha^2}{\gamma H_k})\frac{d \hat{m}}{dt} = -\hat{m} \times \Vec{h}_{eff} - \alpha \hat{m} \times \hat{m} \times \Vec{h}_{eff} \\ - \hat{m} \times \hat{m} \times \Vec{i}_s + \alpha \hat{m} \times \Vec{i}_s
\end{multline}

\begin{figure}[b]
	\centering
	\subfloat[]{\includegraphics[width=0.49\linewidth]{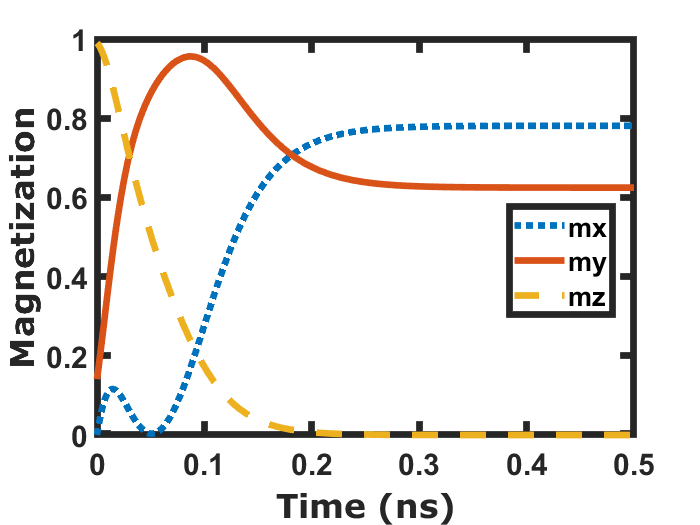}\label{fig:sssdfffsd}}
	\subfloat[]{\includegraphics[width=0.49\linewidth]{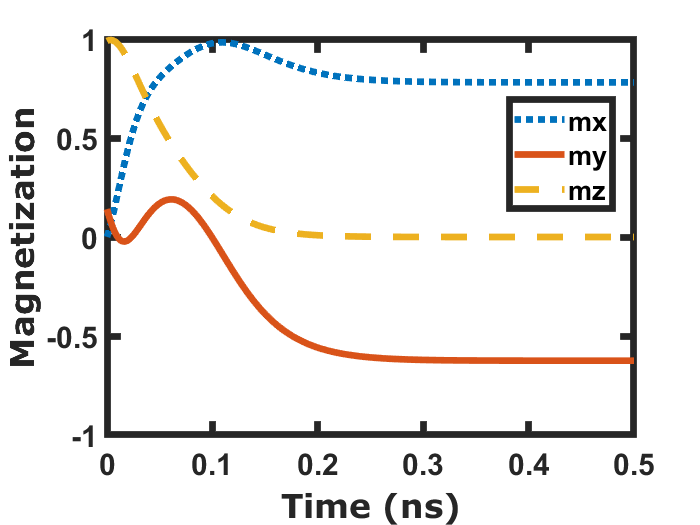}\label{fig:rersdfdf}}
	
	\caption{Magnetization dynamics of the free FM layer for $I_{cy}=340 \mu A$ and (a) $I_{in}=266 \mu A$  (b) $I_{in}=-266 \mu A$ .}
	\label{mag_dyna}
\end{figure}

The spin current can be resolved as
\begin{equation} \label{eq:spin_curr}
    \Vec{i}_s = i_{s,m} \hat{m} + i_{s,\parallel} \hat{N} + i_{s,\perp} \hat{N} \times \hat{m},
\end{equation}
where $\hat{N}$ is the unit vector along some direction,
%where $\hat{N}$ is the unit vector along the direction of the anisotrpy field of the free FM,
using \eqref{eq:llg_main} and \eqref{eq:spin_curr} LLGS equation can be modified as
\begin{multline} \label{eq:llg_mod}
  (\frac{1+\alpha^2}{\gamma H_k})\frac{d \hat{m}}{dt} = -\hat{m} \times \Vec{h}_{eff} - \alpha \hat{m} \times \hat{m} \times \Vec{h}_{eff} \\
  - \hat{m} \times \hat{m} \times (i_{s,\parallel}+ \alpha i_{s,\perp}) \hat{N} \\
  - \hat{m} \times (i_{s,\perp} - \alpha i_{s,\parallel}) \hat{N}\\
\end{multline}
From \eqref{eq:llg_mod} it can be inferred that the spin current $ (i_{s,\parallel}+ \alpha i_{s,\perp})$ (the Slonczewski term) acts as a damping/anti-damping term and $(i_{s,\perp} - \alpha i_{s,\parallel})$ acts as a precession term with field-like behaviour along $\hat{N}$.
To generate linear rotation in magnetization of the free FM and linear resistance change in the MTJ , the magnitude of the field-like term needs to be increased with $\hat{N}$ along $\hat{y}$. A fixed $i_{s,\perp}$ cannot be applied to the FM using the spin hall effect as its magnitude ($\hat{N} \times \hat{m}$) depends on the magnetization of the FM ($\hat{m}$) and changes continuously as the magnetization is rotated. The field-like term can also be changed via $i_{s,\parallel}$ whose direction (along $\hat{N}$) is fixed, but it requires 1/$\alpha$ ($\approx 100$) times more current. And $i_{s,\parallel}$ also affects the magnetization through the damping term. So, this method is not feasible for generating continuous rotation of magnetization of free-FM.

\subsection{Orthogonal spin current driven MTJ}
\label{appendix:c}
Our design uses orthogonal spin currents to generate the continuous rotation of magnetization in the free FM layer. These spin currents are produced by injecting two currents into the heavy metal layer.

Figure~ \ref{mag_dyna} shows the magnetization dynamics of the rotation. The dynamics are examined for $I_{in}=266 \mu A$ and $-266 \mu A$  while the orthogonal current($I_{cy}$) is kept at $340 \mu A$. The settling time measured is $0.2 ns$.

\begin{figure}[]%!htbp
	\centering
    \includegraphics[scale=0.3]{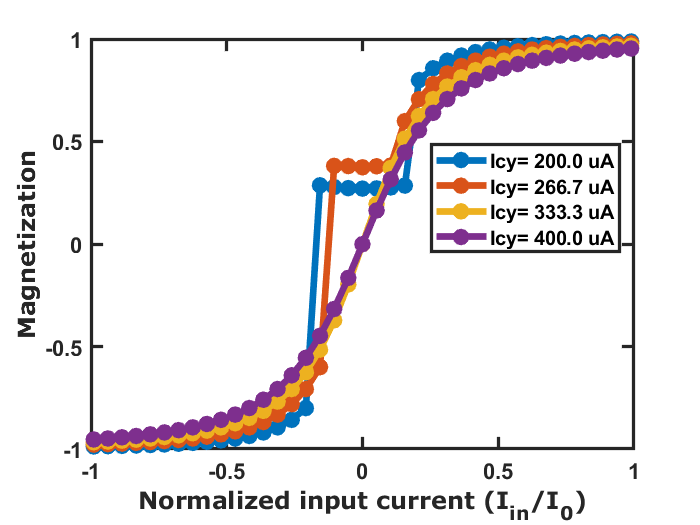}
	\caption{Magnetization($m_y$) of free FM with varying $I_{in}$, for multiple $I_{cy}$. }
	\label{multi_isx}
\end{figure}

One of the orthogonal currents is fixed ($I_{cy}$) to get the continuous rotation in magnetization, while the other current($I_{in}$) is varied. The $I_{cy}$ current needs to cross a threshold to generate the continuous rotation. Fig. \ref{multi_isx} shows that $I_{cy}$ needs to be greater than $334 \mu A$ to achieve the continuous rotation in magnetization of the free FM layer for $\Delta = 45$.

\section{Stability of Magnetization}
\label{appendix:mag_stability}
In this section, we explore different ways to validate the stability of magnetization.

\subsection{Energy profile}
The stability of a system can be usually found from the energy profile.
But, in the presence of the spin current, the energy profile ($E=-\Vec{M}\cdot \Vec{H}_{eff}$) of the free ferromagnet (FM) does not change significantly due to the non-conservative nature of the Slonzewski spin torque \cite{salahuddin2008spin}.   

In the presence of spin current, the LLGS equation is given by 
\begin{multline} \label{eq:llg}
  (\frac{1+\alpha^2}{\gamma })\frac{d \hat{m}}{dt} = -\hat{m} \times (\Vec{H}_{eff} - \alpha \beta \Vec{I}_s ) -  \hat{m} \times \hat{m} \times (\alpha \Vec{H}_{eff} + \beta \Vec{I}_s )
\end{multline}

Here, $\beta= \frac{\hbar}{2qM_sV}$, the spin current $I_{s}$, and $\alpha$ is the damping factor. For small value of $\alpha$,  $\alpha \beta \Vec{I}_s<<\Vec{H}_{eff}$  the LLGS equation can be rewritten as 
\begin{multline} \label{eq:llg}
  (\frac{1+\alpha^2}{\gamma })\frac{d \hat{m}}{dt} = -\hat{m} \times \Vec{H}_{eff} -  \hat{m} \times \hat{m} \times (\alpha \Vec{H}_{eff} + \beta \Vec{I}_s )
\end{multline}
It can be seen from Equation 2 that the spin current in the LLGS equation act either as a damping torque or an anti-damping torque. Being a non-conservative term \cite{salahuddin2008spin}, the spin torque can not affect the energy profile of the FM. One may draw an analogy with the friction of motion, which may either opposes the motion (damping term: always the case for friction) or may hypothetically support the motion (anti-damping), but it can not change the conservation energy profile (e.g., gravitational potential energy) of the system. Hence, we can not employ the energy profile of the ferromagnet in the presence of orthogonal spin currents (in our case) to analyze the stability of the magnetization.
\\
\begin{figure*}[]
\centering
\subfloat[]{\includegraphics[width=0.35\linewidth,height=4.5cm]{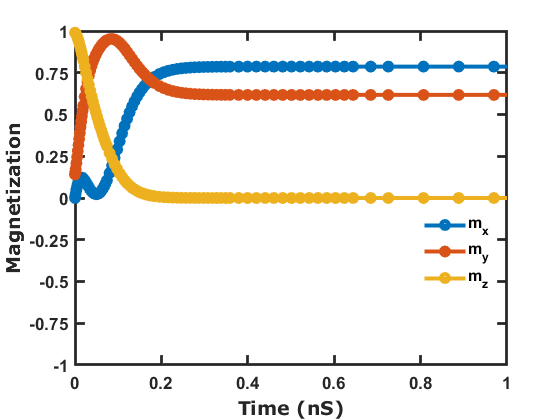}\label{fig_mag_time1}}
\subfloat[]{\includegraphics[width=0.35\linewidth,height=4.5cm]{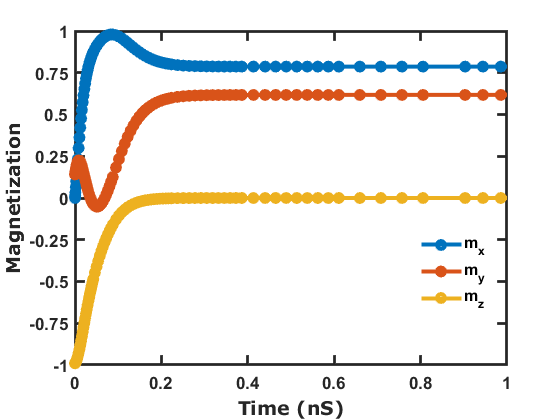}\label{fig_mag_time2}}
	\subfloat[]{\includegraphics[width=0.25\linewidth,height=4.5cm]{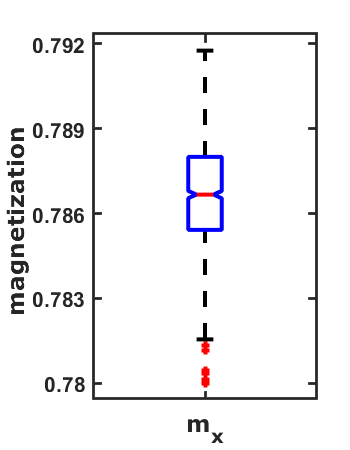}\label{fig:err_mx}}\\
	\subfloat[]{\includegraphics[width=0.25\linewidth,height=4.5cm]{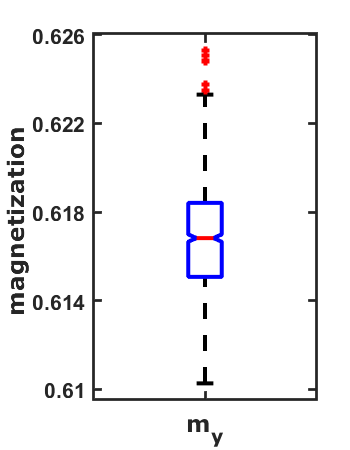}\label{fig:err_my}}
        \subfloat[]{\includegraphics[width=0.25\linewidth,height=4.5cm]{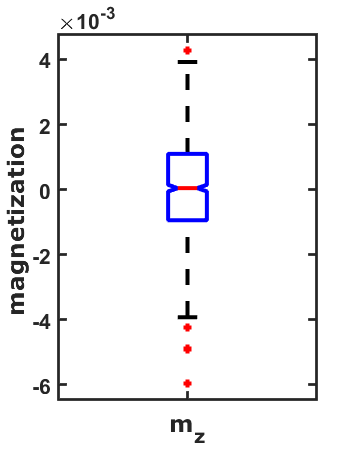}\label{fig:err_mz}}
	\subfloat[]{\includegraphics[width=0.4\linewidth,height=4.5cm]{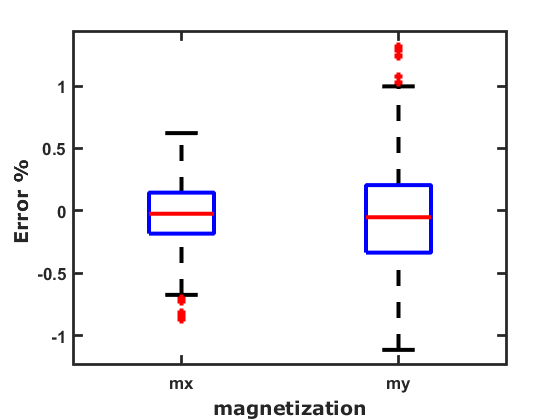}\label{fig:err_per_mx_my}}
\caption{(a), (b) Magnetization dynamics of free-FM of the MTJ with an applied spin current of 510$\mu$A with x-polarization and 400$\mu$A with y-polarization. It can be seen here that the magnetization is stable after it reaches its final value of $m_x=0.7869, m_y=0.6171, m_z=0.00$ with different initial magnetizations of -z and +z. (c), (d), (e) and (f) Noise analysis of magnetization dynamics of the free FM with 1000 simulations of LLGS equation under the influence of thermal noise, having a thermal stability factor($\Delta$) of $45$. (c) Variation in $m_x$ magnetization due to thermal noise. (d) Variation in $m_y$ magnetization due to thermal noise. (e) Variation in $m_z$ magnetization due to thermal noise. (f) Boxplot of error $\%$ of $m_x$ and $m_y$ magnetization.}
\label{fig_ma}
\end{figure*}

\subsection{LLG simulation}
The stability of the magnetization in the presence of the spin current can be analyzed via evaluation of the magnetization from the different initial configurations and/or stochastic evaluation of the magnetization with thermal noise ($ \langle H_{th}^2 \rangle = \frac{2\alpha k_B T}{\gamma M_s V}$). We show in Fig. \ref{fig_mag_time1}\&(b) the magnetization dynamics of the FM in the presence of orthogonal spin currents 510$\mu$A, 400$\mu$A with x-and y-polarizations respectively, producing a stable magnetization orientation of $m_x=0.7869, m_y=0.6171, m_z=0.00$. It can be observed (Fig. \ref{fig_mag_time1}\&(b)) the different initial magnetization of the FM stabilizes to the same final value for given spin current inputs. Figures \ref{fig:err_mx},\ref{fig:err_my} and \ref{fig:err_mz} show the effect of thermal noise on $m_x, m_y$, and $m_z$ magnetizations respectively. The error percentage of magnetization is shown in Fig. \ref{fig:err_per_mx_my}, it can be inferred that the thermal noise has minimal effect on the magnetization with error $\leq 1\%$,  leading to a stable magnetization rotation in the presence of the spin currents. The error is estimated by performing 1000 stochastic simulations.\\

\subsection{Net torque on the FM}

The stability of magnetization can be analyzed from the plots of torque vs polar angle ($\theta$) and azimuthal angle ($\phi$) of the magnetization.

\begin{figure*}[]
	\centering

	\subfloat[]{\includegraphics[width=0.3\linewidth, height=4cm]{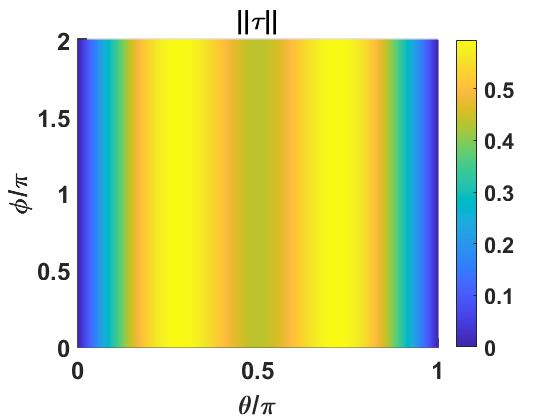}\label{fig:us_swt_t_total}}
	\subfloat[]{\includegraphics[width=0.3\linewidth, height=4cm]{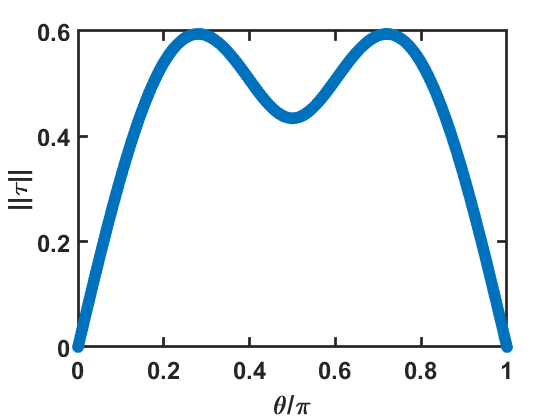}\label{fig:us_swt_t_total_phi_0}}
 \subfloat[]{\includegraphics[width=0.3\linewidth, height=4cm]{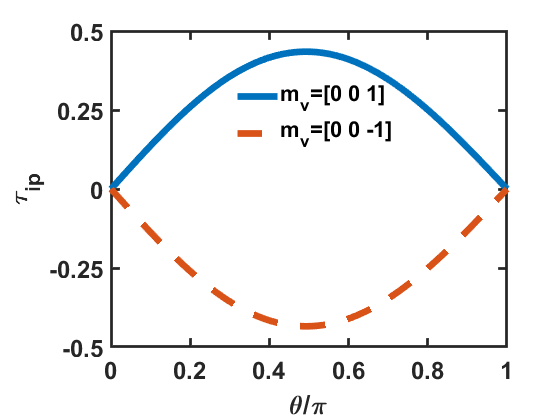}\label{fig:usu_swit_t_in_plane}}
	\caption{(a) Norm of the net torque acting on the free-FM in the presence of the spin current of $500\mu A$ with z-polarization. The net torque vanishes at $\theta=0$ and at $\theta=\pi$. (b) The norm of the net torque with $\theta$ at $\phi=0$ is plotted to clearly spot the vanishing torque points. (c) The in-plane torque with $\theta$ for the vanishing torque points at $\phi=0$.}
	\label{meh}
\end{figure*}

\begin{figure*}[]
	\centering

	\subfloat[]{\includegraphics[width=0.3\linewidth, height=4cm]{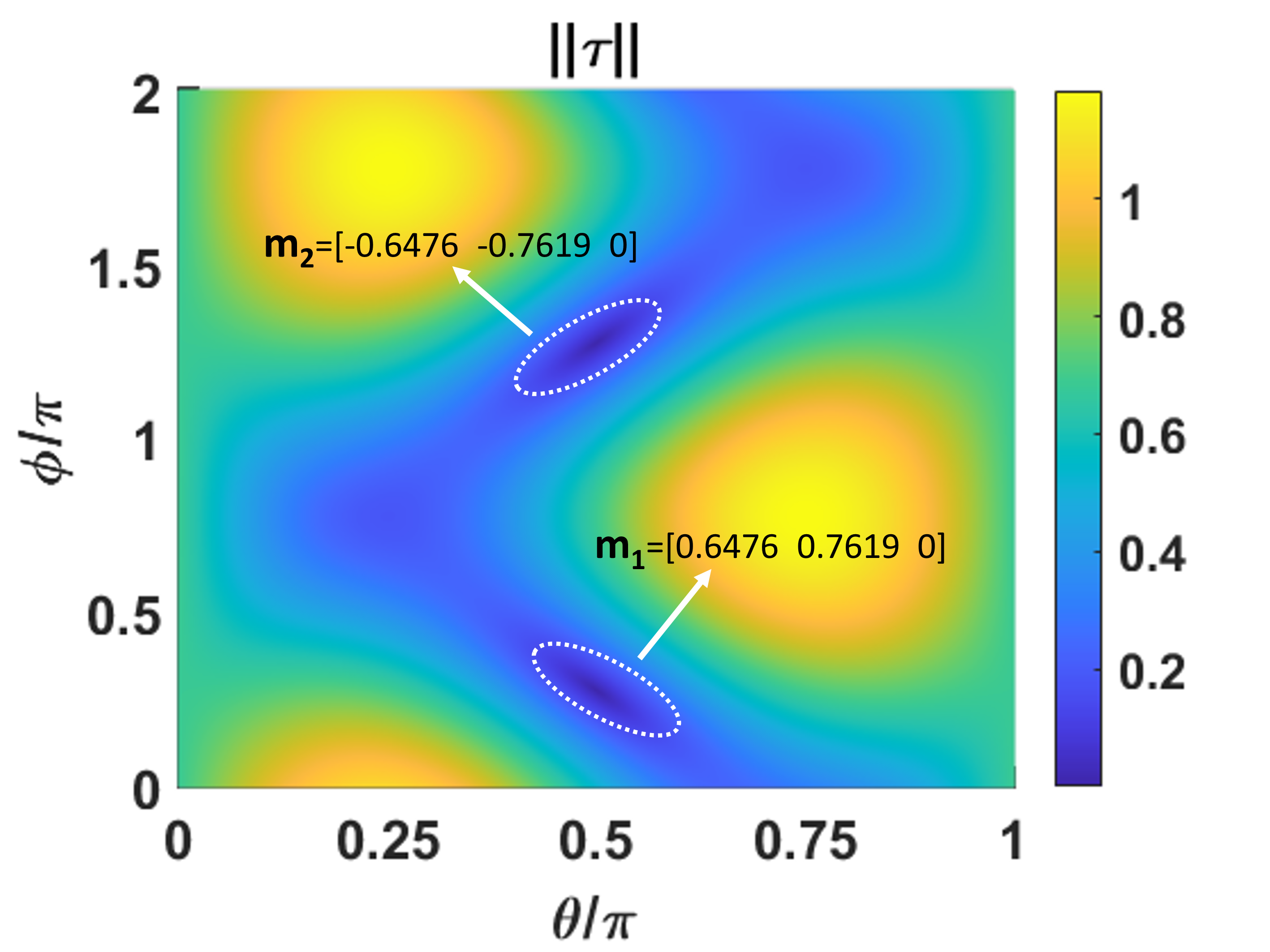}\label{fig:ortho_swit_t_total}}
	\subfloat[]{\includegraphics[width=0.3\linewidth, height=4cm]{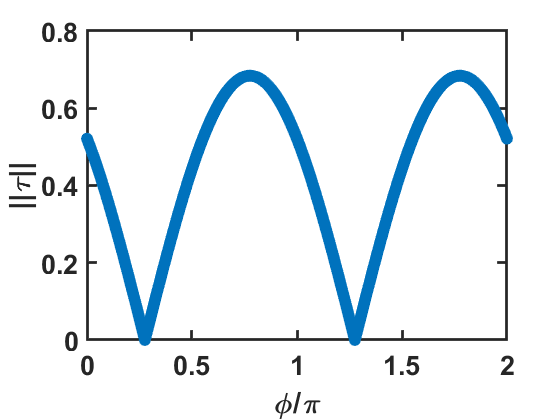}\label{fig:ortho_swit_t_total_theta_1.57}}
 \subfloat[]{\includegraphics[width=0.3\linewidth, height=4cm]{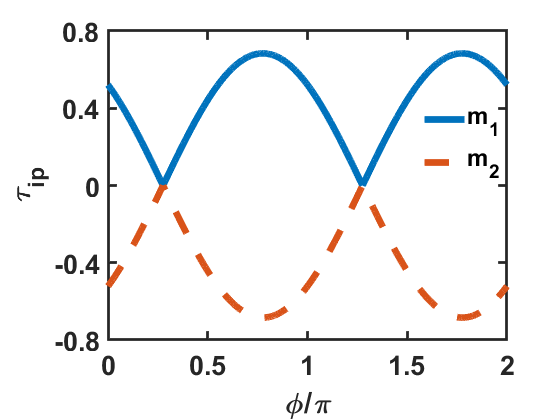}\label{fig:ortho_swit_t_in_plane}}
	\caption{(a) Norm of the net torque acting on the free-FM in the presence of the spin current of $510\mu A$ and $600\mu A$ with x \& y-polarization, respectively. The net torque vanishes at $m_1$ and $m_2$ points. (b) The norm of the net torque with $\phi$ at $\theta=\pi/2$ is plotted to clearly spot the vanishing torque points. (c) The in-plane torque with $\phi$ for the vanishing torque points at $\theta=\pi/2$.}
	\label{meh}
\end{figure*}
The net torque on the free FM is given by:
\begin{equation}
\label{eq:}
  \Vec{\tau} = -\hat{m} \times \Vec{H}_{eff} - \alpha \hat{m} \times \hat{m} \times \Vec{H}_{eff}  - \beta \hat{m} \times \hat{m} \times \Vec{I}_s + \alpha \beta \hat{m} \times \Vec{I}_s
\end{equation}
The net torque($\tau$) can be decomposed into in-plane torque($\tau_{ip}$) and out-of-plane torque ($\tau_{op}$). The $\tau_{ip}$ can be damping (positive) or anti-damping (negative). The $\tau_{ip}$ is given by
\begin{equation}
    \tau_{ip}= \hat{i}_p \cdot \Vec{\tau}
\end{equation}
where, $\hat{i}_p$ is given by
\begin{equation}
    \hat{i}_p= \frac{-\hat{m} \times \hat{m} \times \hat{m}_{v}}{|-\hat{m} \times \hat{m} \times \hat{m}_{v}|}
    \\
\end{equation}
Here, $\hat{m}_{v}$ is the vanishing torque point($\Vec{\tau}=0$)\cite{roy2012metastable}, which can be extracted from plots of torque vs  $\theta$ and $\phi$ of the magnetization. If the $\tau_{ip}$ is damping(positive), that means the $\hat{m}_{v}$ is a stable point, and if it is anti-damping(negative), then $\hat{m}_{v}$ is not a stable point in the presence of the applied spin current.  We first demonstrate the effectiveness of this method on regular switching(anti-parallel to parallel, i.e., PMA injected with z-polarized spin current or vice versa) and adapt the same to explain the magnetization rotation via orthogonal spin current injection considered in the paper.

\subsection*{Regular switching}
Consider a PMA ferromagnet injected with a spin current of $500\mu A$ with z-polarization, higher than the critical spin current for switching. It is well known that the resulting magnetization will be aligned in the +z-direction. So, in the presence of spin current, +z is the stable point, whereas -z is not a stable point.

Figure \ref{fig:us_swt_t_total} shows the norm of the total torque acting on the free FM. It shows two regions ($\theta=0,\pi$ pointing to +z and -z directions, respectively) where the torque becomes zero which can be clearly seen from Fig. \ref{fig:us_swt_t_total_phi_0}. So, even in the presence of spin current, the net torque vanishes at the -z direction, this means that the net torque alone cannot fully explain the stability of the magnetization.

To fully verify the stability of the magnetization point, we can look at the in-plane torque($\tau_{ip}$). Figure \ref{fig:usu_swit_t_in_plane} shows the in-plane torque for both of the vanishing torque points. It can see that for $\hat{m}_{v}=$[0 0 1], the $\tau_{ip}$ is always positive(damping), implying a stable magnetization point, whereas for $\hat{m}_{v}=$[0 0 -1], the $\tau_{ip}$ is always negative(anti-damping) implying not a stable magnetization point in the presence of the spin current.

\subsection*{Orthogonal spin current injected free FM}
The stability of the magnetization considered in this paper can be explored in a similar fashion. The spin currents of $510\mu A$ and $600\mu A$ with x \& y-polarization, respectively, are injected into the PMA free-FM, and the expected stable point for this condition is $m=$[0.6477 0.7619 0]. We show in Fig. \ref{fig:ortho_swit_t_total} the net torque on the free-FM. It can be seen that net torque vanishes at two points($m_1=$[0.6476 0.7619 0] and $m_2=$[-0.6476 -0.7619 0]), and both the torque vanishing points appear at $\theta=\pi/2$, these points can be clearly seen from Fig. \ref{fig:ortho_swit_t_total_theta_1.57}.

We show the in-plane torque in Fig. \ref{fig:ortho_swit_t_in_plane} to inspect the stability of these torque vanishing points. The $\tau_{ip}$ is always positive for $m_1$ implying the stable point, and always negative for $m_2$, implying not a stable point in the presence of the orthogonal spin currents. We believe that the thermal stability of the magnetization may be correlated to the area under the in-plane torque curve, which will be addressed in our future works.

\bibliographystyle{IEEEtran}
\bibliography{reference}

\end{document}